\documentclass[]{spie}  

\usepackage{amsmath,amsfonts,amssymb}
\usepackage[utf8]{inputenc}
\usepackage{graphicx}
\usepackage[colorlinks=true, allcolors=blue]{hyperref}
\usepackage{cleveref}
\usepackage{subcaption}
\usepackage{bookmark}
\usepackage{blindtext}
\usepackage{physics}
\usepackage[version=4]{mhchem}
\newcommand{\mean}[1]{\langle {#1}\rangle}
\newcommand{\comment}[1]{}

\title{A NICE (Nulling Interferometry Cryogenic Experiment) update: Beyond 1e-5 and towards cryogenic operation}

\author[a]{Jonah T. Hansen}
\author[a]{Thomas Birbacher}
\author[a]{Germain Garreau}
\author[a]{Julio Pino-Jiménez}
\author[a]{Emilie Bouzerand}
\author[a]{Felix A. Dannert}
\author[a,b]{David Eglin}
\author[a]{Lee Feinberg}
\author[a]{Adrian M. Glauser}
\author[a]{Maximilian Kirchhoff}
\author[b]{Oliver Pitz}
\author[a]{Eckhart Spalding}
\author[a,c]{Sascha P. Quanz}
\affil[a]{ETH Zurich, Institute for Particle Physics and Astrophysics, Wolfgang-Pauli-Str. 27, 8093 Zurich, Switzerland}
\affil[b]{ETH Zurich, Institute for Quantum Electronics, Auguste-Piccard-Hof 1, 8093 Zurich, Switzerland}
\affil[c]{ETH Zurich, Department of Earth Sciences, Sonneggstrasse 5, 8092 Zurich, Switzerland}

\authorinfo{Send correspondence to J.T.H, E-mail: johansen@phys.ethz.ch}

\begin{document} 
\maketitle

\begin{abstract}
The Nulling Interferometry Cryogenic Experiment (NICE) is a mid-infrared laboratory testbed at ETH Zürich that aims to not only reproduce the deep ($<1\times10^{-5}$) broadband nulls of the nulling testbeds of the early 2000s, but also at the required sensitivity levels expected for the LIFE space mission. This sensitivity enforces the experiment to go cryogenic at temperatures around 15\,K; an ambitious task for an ultra-precise interferometer. We share our results of the ambient precursor experiment, demonstrating repeatable $<1\times10^{-5}$ nulls at a single wavelength and high throughput, and investigations into the nulling performance across a broader bandpass and with dual polarisation states. We will also highlight the push towards cryogenic operation, with the instalment of a new 15\,K test cryostat that will inform our choices of materials and optomechanical mounting techniques.

\end{abstract}


\section{INTRODUCTION}
\label{sec:intro}  

The Nulling Interferometry Cryogenic Experiment (NICE) is one of the key developments in the development of the Large Interferometer For Exoplanets (LIFE) Space Mission \cite{Quanz-2022-ID17,Glauser-2024-ID1} -- a mid-infrared (MIR) formation-flying nulling space interferometer designed to detect and characterise potentially habitable exoplanets around the solar neighbourhood. Such a mission requires demonstration of its instrument and measurement concept, which is the purpose of the NICE testbench: to mature the optical and mechanical concept through experimentation at both ambient and cryogenic conditions.
The ultimate aim is to reproduce the deep $(<1\times10^{-5})$, broadband ($>20$\% bandwidth) nulls of earlier nulling testbeds \cite{Peters_2008,Gappinger_2009,Martin_2012,LeDuigou_2012}, while simultaneously measuring a planet signal at the required sensitivity for LIFE - necessitating the instrument to be cooled to temperatures around 15\,K. 

To date, NICE has primarily focused on the ambient precursor bench - the main purpose of which is to ensure the concept works before implementing the full system at cryogenic temperatures. The paper by Birbacher \& Hansen et al. (2026) \cite{Birbacher-Hansen-2026}, along with previous proceedings of this conference \cite{Ranganathan-2024-ID9,Birbacher_2024}, describes some of the key results of the early nulling tests, primary of which was the demonstration of a $7.2\times10^{-6}$ null at 4.7\,\textmu m over 25\,s, with a corresponding 17\% throughput. We point the reader to this paper for more information on the requirements of the experiment and the optical design.

In this proceeding we will summarise the current status of the project, which has been approached from three different directions: continued progress on the ambient bench, optical simulations to inform manufacturing and alignment tolerances, and ongoing work to prepare the system for cryogenic operation. We will discuss each in turn, before outlining the future roadmap for the experiment. More detailed information on some of these developments can be found in other proceedings of this conference, which will be outlined where appropriate. 

\section{NICE optical summary}
\label{sec:summary}

\begin{figure}
    \centering
    \includegraphics[width=0.8\linewidth]{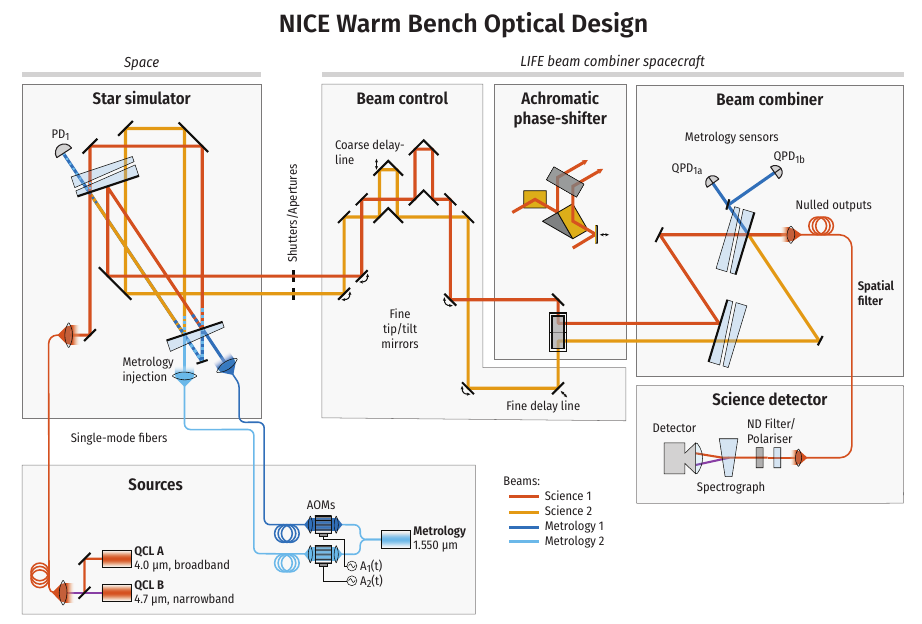}
    \caption{Optical diagram of the NICE ambient optical table. Adapted from Birbacher \& Hansen et al. (2026) \cite{Birbacher-Hansen-2026}.}
    \label{fig:optical_schematic}
\end{figure}

Before describing the updates and challenges NICE has faced in the recent months, we will provide a quick summary of the current ambient optical bench. As previously mentioned, we point the reader to Birbacher \& Hansen et al. (2026) \cite{Birbacher-Hansen-2026} for more details. The optical schematic is shown in \cref{fig:optical_schematic}. The bench is currently a two-beam Bracewell interferometer, and as such does not represent the full four-beam LIFE instrument complete with cross-combination and phase chopping.

The bench has two science quantum cascade lasers: one narrowband distributed feedback laser at 4.7\,\textmu m, and a broadband ($\sim 3\%$) Fabry-Perot laser at 4.0\,\textmu m. These are located on a separate table, and are carried over to the main table via an \ce{InF_3} single mode fibre (SMF). The beam then enters the ``star simulator'', a modified Sagnac interferometer, which splits the beam into two with ideally equal amplitude and polarisation. This subsystem is not part of the LIFE instrument, but necessary to simulate the astrophysical scene. 

A 1550\,nm metrology laser is injected via a beam splitter and co-aligned both forwards and backwards with the science beam. Through measurements at a rear photodiode and two quad-cell photodiodes at the end of the optical train (see \cref{fig:optical_schematic}), the optical path difference (OPD), pointing and lateral position of the metrology (and by proxy, science) beams can be measured. More on this measurement technique can be found in Birbacher et al. (2024) \cite{Birbacher_2024}. 

After the star simulator, the two science beams then enter the ``combiner'' spacecraft, whereby we use a number of fine tip/tilt mirrors, a coarse delay line (20\,nm precision) and a fine delay line ($<1\,$\,nm precision) to correct for alignment and phase. The tip/tilt stages and fine delay line are run in closed loop with the metrology sensors to ensure a stable setup. To implement destructive interference for an on-axis source over the whole wavelength range, we implement a achromatic phase shifter in the form of a reflective periscope to flip the polarisation states. 

The now corrected beams are then combined in a modified Mach Zehnder (MMZ)\cite{Serabyn_2001} combiner, to ensure each beam undergoes the same number of reflections and transmissions. This combiner architecture produces two nulled outputs; however, due to the need to duplicate the control optics inside the beam combiner to correct the second output, we currently only analyse the null from one of them. This results in a loss of 50\%, but is one of the foreseen upgrades to the bench in the near future. 

Finally, the combined beams are injected into another \ce{InF_3} SMF. This is a crucial part of the setup, allowing any phase variations across the pupil to be rejected by the fibre and resulting in a more easily managed intensity loss. The output light is then collimated and sent through a polariser, ND filters and a dispersive prism before landing on an \ce{InSb} detector. 

\begin{figure}
    \centering
    \includegraphics[width=0.8\linewidth]{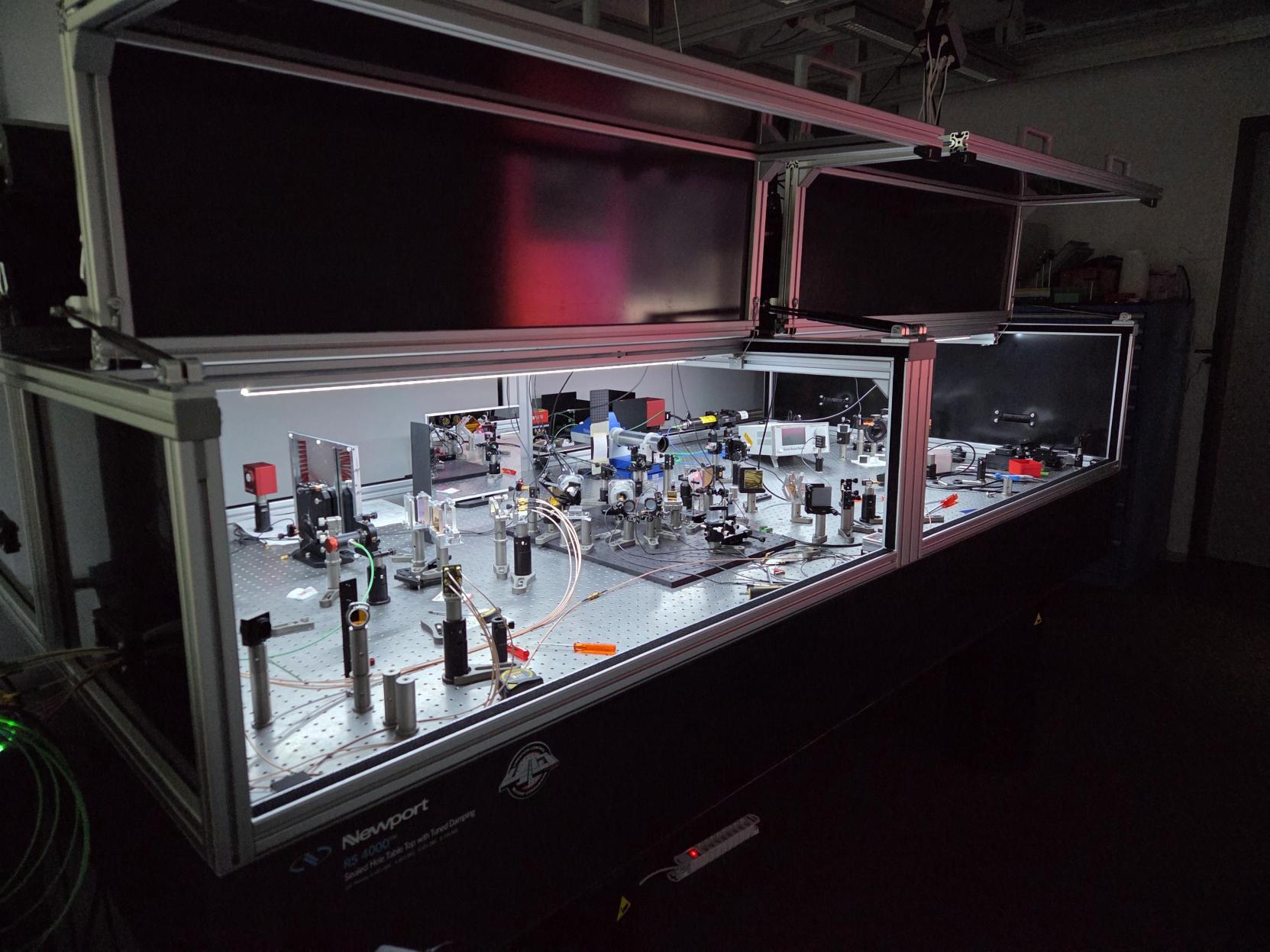}
    \caption{Photograph of the NICE ambient optical table}
    \label{fig:bench_photo}
\end{figure}

We note here that the bench has undergone several major changes in mounting and alignment since the previous proceedings, such as a large shrinkage in optical footprint, as well as moving from kinematic mounts to static posts to assist with stability. The bench has also recently obtained a new enclosure to help reduce the effects of air turbulence - a particular concern due to the lab's placement on the top floor of a high building on the ETH campus. A photo of the bench with the enclosure is shown in \cref{fig:bench_photo}

\section{Optical model and tolerancing}
\label{sec:tolerancing}

\begin{figure}
    \centering
    \includegraphics[width=0.8\linewidth]{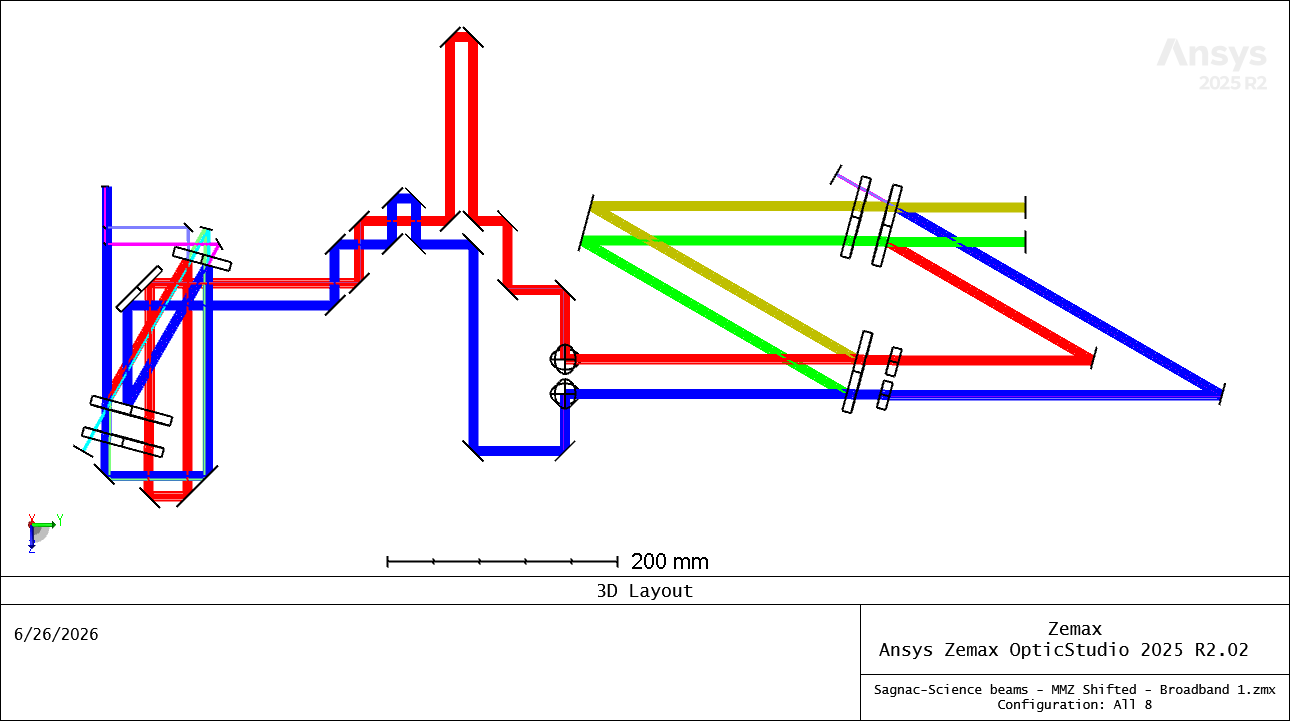}
    \caption{Zemax optical model of the NICE ambient bench, containing both the science and metrology beams}
    \label{fig:optical model}
\end{figure}

In order to derive and verify the opto-mechanical tolerances needed for the experiment, particularly ensuring alignment at cryogenic conditions, we have implemented an optical model in Zemax; this model is shown in \cref{fig:optical model}. It should be noted that this model includes components not currently in the ambient bench - namely the two outputs of the MMZ, and two longitudinal dispersion correctors (found after the first beamsplitter in the MMZ). Here, we use a merit function that indirectly acts on the null depth, given by:
\begin{equation}
    \label{eq:nice1_null-dynamic}
    \mean{N}
    = \frac{1}{4} \Big( \mean{\delta\phi}^2 + \frac{1}{4}\mean{\delta\phi_\text{sp}}^2 + \mean{\delta I}^2 + \frac{1}{4} \mean{\delta I_\text{sp}}^2 \Big),
\end{equation}
where $\mean{N}$ is the mean null depth at a single wavelength and spatial mode, $\delta\phi$ is the phase error between the beams, $\delta I = \abs{I_1 - I_2}/(I_1 + I_2)$ is the intensity mismatch between beams, and terms with an $sp$-subscript indicate differential errors between the $s$ and $p$ polarisation modes. At 4\,\textmu m, from Birbacher \& Hansen et al. (2026)\cite{Birbacher-Hansen-2026}, we have the following requirements for the terms in \cref{eq:nice1_null-dynamic}:
\begin{itemize}
    \item $\langle \delta\phi \rangle \leq 2.3\,$nm,
    \item $\langle\delta\phi_{sp}\rangle \leq 4.7\,$nm,
    \item $\langle \delta I \rangle \leq 0.3\,$\%,
    \item $\langle\delta I_{sp}\rangle \leq 0.74\,\%$.
\end{itemize}
We only consider static errors, as dynamic errors are governed by the closed loop control system rather than the opto-mechanical mounting scheme. High-order wavefront error is also excluded, as they require surface figure modelling that is not currently included in the model. 

One of the current intentions is to determine whether we can manufacture each subsystem as a single enclosure with all optics integrated. Here, each subsystem can be manufactured and tested separately and, upon meeting the specifications, they can be aligned relative to one another. We also include a range of compensator mechanisms, namely the linear stages and tip/tilt actuators shown in \cref{fig:optical_schematic}, to correct the perturbed system. For our initial results, we assume:
\begin{itemize}
    \item $\pm$5\,\textmu m machining tolerances, given by the machine used at the Physics workshop at ETH Zurich
    \item $\pm$15\,\textmu m decentre and tilt tolerances of each subsystem alignment, accounting for accumulation of error across each subsystem.
    \item Stroke and precision of off-the-shelf actuators from suppliers such as JPE and Physik Instrumente. 
    \item Wavelength bandpass of 4-8\,\textmu m, corresponding with the transmission window of \ce{CaF_2}
\end{itemize}

A Monte-Carlo simulation was selected to assess the suitability of the tolerances assumed above. Within this framework, we perturb every individual optical element and subsystem interface simultaneously by the assumed tolerances, after which the compensators engage to correct the system. The actuator itself is then perturbed further to replicate the finite precision of the mechanism. The compensation procedure directly emulates the alignment procedure of the bench to ensure that the correct degrees of freedom are compensated for, and such that there is a clear observable for the procedure at each given step (namely either null depth, or intensity through the fibre). 

Overall, under the above assumptions, we find that the system has a highly robust response against perturbations, with a compliance yield of 95\% across most key performance parameters over the full bandpass. A notable exception is of the intensity mismatch which has a 75\% compliance. This is tied to the coatings on the mirrors and especially the beam splitters, whereby if a misalignment breaks symmetry (even if compensated), the polarisation and incident angle dependence on the R:T ratio of the beam splitter will result in a mismatch. Such interdependent errors and cross-terms are critical to find and mitigate.

While this model accounts for a wide array of system variables, there are a number of limitations and ``blind spots'' that will be addressed in future stages of this work. They include:
\begin{itemize}
    \item Ideal optical components: the system does not consider for surface figure errors and material inhomogeneities. Such errors are likely to break the symmetry of the system and drive many of the key tolerances.
    \item Angle-dependent coatings: the current coatings assume a contrast efficiency as a function of angle, the true nature of which has not yet been measured. This will likely accentuate the coating errors described above. 
    \item Thermal drift in alignment or material properties due to thermal gradients have currently been excluded.
    \item No scattering or stray-light is assumed in the model.
    \item The simulation assumes a simplified geometric approximation to approximate the coupling into the SMF. For a realistic system, especially to account for high-order mode leakage, physical optical modelling of the fibre is required.
\end{itemize}

Should the above effects result in non-compliance of the optical model, we will need to then either impose even stricter mechanical tolerances (driving up cost) or consider more exotic compensation mechanisms (see  Birbacher et al., proc. 14148-119 \cite{Birbacher_SPIE_2026}). 

\section{Ambient bench activities}
\label{sec:ambient}

Since obtaining the deep, single wavelength and single polarisation null reported in Birbacher \& Hansen et al. (2026) \cite{Birbacher-Hansen-2026}, a lot of work has gone into trying to stretch the abilities of the nuller in three main directions: repeatability \& stability through better spatial filtering, broadband nulling, and polarisation agnostic nulling. We will describe updates, key blockers and areas for future study in turn.

\subsection{Investigation into improved spatial filtering}
\label{sec:spatial_filtering}

Spatial filtering via an SMF is arguably one of the most important parts of the nuller, whereby phase variations across the pupil turn into bulk intensity losses; greatly reducing the requirements on wavefront error at the expense of throughput. However, this relies on a key premise: that light not belonging to the fundamental mode of the fibre does not make it through to the output. Through testing, we have found that this is not the case, and that some high-order modes make it through the cladding. 

This problem is not new, being characterised by some of the early efforts into the TPF-I and \textit{Darwin} missions \cite{Ksendzov2007,Cheng2009}. Such an effect is subtle, as it only appears when we null deeper than $1\times10^{-4}$ -- manifesting as a null floor with unexpected spatial structure. This is a variable floor, as it strongly depends on the injection; the stronger the misalignment, the more mode leakage is seen. Due to vibrations and perturbations in the earlier optics, the variable injection leads to a non-repeatable, unstable setup. Therefore, this indicates that our \ce{InF_3} fibre, despite being 5\,m long, does not have the required suppression of high order modes for our current alignment tolerances. 

To combat this issue, we have tried a variety of fixes. Firstly, we tried new fibre materials including chalcogenide glass (specifically \ce{As_2S_3}). As was expected, however, these fibres had substantially less throughput (relative core throughput loss of 16\,dB compared to a comparable \ce{InF_3} fibre). We also found that the high-order cladding modes were not-suppressed at all; while the core was clean, the comparably low acrylate coating on the fibre resulted in negligible cladding attenuation -- essentially acting as a high NA multimode fibre. We are in discussions to procure a variant of the fibre with a liquid gallium coating to efficiently strip out these modes \cite{Ksendzov2007,Cheng2009}; this will be tested for its efficiency and its ability to function at cryogenic temperatures. A shorter fibre (to counteract the throughput loss) with an efficient coating holds promise due to it much wider wavelength range.

In a similar vein, we tried changing the coating of our \ce{InF_3} fibre. This was done by stripping the acrylate coating, applying a layer of high index epoxy, and then connectorising the fibre. Regrettably, somewhere during this process, the fibre must have broken (possibly near the stripped section), as next to no throughput was able to be recorded. Repeating this procedure on a smaller section of fibre, to ensure that it is not the stripper itself causing issues, may also be investigated.

We also tried masking the output of the fibre with a pinhole, blocking any of the light that would otherwise exit out of the cladding. This technique did appear to work, dropping the amount of high-order mode leakage by about a factor of two. This was not enough to fully remove the effect at the level of the null, and also resulted in a much worse PSF at the detector due to using a \ce{CaF_2} lens over a reflective OAP. Nevertheless, designing output injection optics that support both the pinhole and an OAP would be beneficial. The technique is also likely to be highly effective with the \ce{As_2S_3} fibre. 

To summarise, none of our attempts to suppress the high-order mode leakage of the SMF were successful, but they have led to new ideas and collaborations that may result in better fibres overall. We are also keeping a close eye on endlessly single-mode photonic crystal fibre developments (such as that of Ireland et al., 2024 \cite{Ireland_2024}) which also hold promise for being highly efficient and with a wide spectral bandpass. Finally, we draw the reader's attention to the proceeding of Garreau et al. (proc. 14148-120 \cite{Garreau_SPIE_2026}) that simulates the use of PIAA optics\cite{Guyon2003,Jovanovic2017,Ireland_2024} to substantially increase coupling into fibres over a wide bandpass.

One last ``solution'' to mention is to reduce the amount of unwanted modes in the first place with a more precise and repeatable alignment procedure. This requires a few additional controllable stages (which have recently been procured), and an automatic alignment algorithm. While this will be implemented in the near future, it nevertheless does not fix the underlying problem and reduces the allowed tolerances on injection and intensity balancing through intentional misalignment.

\subsection{Chromatic nulling}
\label{sec:chromatic}

Arguably one of the hardest problems facing NICE, and the LIFE combiner as a whole, is that of the extremely large wavelength range from 4 to 18.5\,\textmu m. In order to sustain an achromatic null, the phase and amplitude of all wavelengths must be matched simultaneously; which only can occur when the path lengths are matched exactly. There are two types of errors we consider: chromatic phase, and chromatic amplitude errors.

To start with the latter, this is where the intensities of one arm may be balanced at one wavelength, but are unbalanced at another. For NICE, we try to minimise this error through the enforcement of symmetry, especially in the MMZ. Each beam undergoes the same number of reflections and transmissions at nominally the same angle of incidence; hence if the two beam splitters and coatings are identical and homogenous, the amplitudes should match as a function of wavelength.

Of course, this is not the case in general as the manufacturing of the coatings and optical surfaces will only be identical to within some tolerance. Our baseline reflective coating is unprotected gold, which has minimal spectral dependence, but contaminants and surface error will result in wavelength dependent wavefront error and, through a mismatch in coupling into the SMF, an imbalance of intensities. The (likely dielectric) beam splitter coatings are even more complex, as they couple polarisation, angle of incidence and wavelength errors together. 

The setup does not currently have a method of compensating chromatic intensity imbalance outside of the $1/\lambda$ dependence of tip/tilt injection into the fibre and a (mostly) achromatic intensity adjustment using a thin wire to vignette the beam. However, upon first measurements with our two lasers at a single polarisation, we have found that the two chromatic intensities are nevertheless the same to within the same error as the summed broadband intensity. This provides strong evidence that the experiment should be able to achieve a broadband null, at least over the 15\% bandwidth between our 4\,\textmu m and 4.7\,\textmu m lasers, and especially once custom (rather than off-the-shelf) optics are used. For LIFE as a whole though, there is still concern that the lack of a dedicated compensator could lead to situations where, once symmetry is broken via something akin to a micrometeoroid strike, the chromatic intensity cannot be balanced within specifications. Furthermore, the extremely large wavelength band makes it very challenging to stay within specifications to begin with.

There are two ways to deal with said error: either splitting the wavelength range into multiple sub-bands that result in chromatic mismatch within requirements, or using an active compensator that can act on chromatic intensity. One such device is known as the ``adaptive nuller'' \cite{Lay_2003,Peters_2008}, using a combination of a dispersive prism and a deformable mirror (DM) to act on phase and intensity as a function of wavelength. This device has been used to obtain the broadest, ``deep'' null on record ($1\times 10^{-5}$ at 20\%)\cite{Peters_2008}, but suffers from the need of a space-qualified, cryogenic DM. More discussion on the trade-off of compensator mechanisms can be found in the proceeding of Birbacher et al. (proc. 14148-119 \cite{Birbacher_SPIE_2026}). 

The second error is that of chromatic phase, sometimes referred to in the literature as longitudinal chromatic dispersion \cite{Leveque_1996,Pannetier_2021}. This is where there is a difference in optical path length as a function of wavelength, which can occur whenever two mediums of different refractive indices are used. Hence, we require not only that the OPD is matched to within requirements (for NICE on the order of 2\,nm), but also that the refractive components have the same thicknesses for each beam. To first order, for a glass prism of refractive index $n(\lambda)$, the thickness variation ($\Delta W$) allowed for a given OPD residual is (assuming an on-axis beam):
\begin{equation}
    \Delta W(\lambda_1,\lambda_2) \leq \frac{2\Delta\text{OPD}}{n(\lambda_1)-n(\lambda_2)}
\end{equation}
where $\lambda_1,\lambda_2$ define the edges of the bandpass. Note that a non-zero incidence angle changes the equation, but not in a significant manner. For a \ce{CaF_2} beam splitter over a wavelength range of 4 to 5\,\textmu m and to achieve an error less than 1\,nm, this is an allowable thickness error of $\sim\pm\,200$\,nm -- a very challenging requirement for manufacturing. 

To compensate for this, we can add a small wedged plate (known as a longitudinal dispersion compensator or LDC) to adjust the amount of glass seen by the beam. In the design of NICE, we include wedged beam splitters to account for ghosting in the combiner, and so are required to also include a wedged compensator to remove the lateral dispersion. This same compensator element can be used as the LDC if translated along the axis of the wedge. To accomplish this, we have developed an integrated mount with an actuated linear stage that can achieve approximately 20\,nm translational precision (for a 0.5 degree wedge, a precision of 170\,pm of glass thickness) with about 5\,mm of stroke.

To date, we have not yet achieved a broadband null via this LDC below approximately 1$\times 10^{-3}$. One potential reason is that we are still considerably far away in stroke. From utilising a CMM machine, we have measured our beam splitters and compensator pairs to have a thickness standard deviation of about 40\,\textmu m, with some pairs reaching differences of nearly 150\,\textmu m. Our current LDC can only correct up to a 40\,\textmu m error, and if one also factors in inhomogeneities across the aperture, there may be additional error coming from the star simulator beam splitter optic. 

As an immediate next step, we plan on integrating a temporary grating at the output to obtain a much higher spectral resolution. With a large scan of OPD, we can then employ a Fourier transform to identify the phase slope and calculate exactly the stroke needed to compensate. At the same time, a higher stroke LDC will also be looked into.

\subsection{Polarisation agnostic nulling}
\label{sec:polarisation}

Since the light emitted by stars and planets is in general  unpolarised, the goal for the LIFE mission is to achieve a null depth that is polarisation agnostic. From \cref{eq:nice1_null-dynamic}, we can see that this means the polarisation must be matched carefully between all beams, and thus is another driver of preserving symmetry in the instrument. Several elements may impact the differential polarisation state between our two beams: the coating of our refractive optics, strong misalignment of reflective elements, or polarisation scrambling in the fibre. 

\subsubsection{Characterisation technque}
The first step is to characterise the polarisation state of the two beams in NICE, especially from the star simulator to the end of the beam combiner where differential effects are likely to arise. The sources that we use are polarised, and MIR depolarisers are not available; as such, we must perform analysis on the two orthogonal input polarisation states and ensure that the nulling performance holds for both simultaneously.

\begin{figure}
    \centering
    \includegraphics[width=0.8\linewidth]{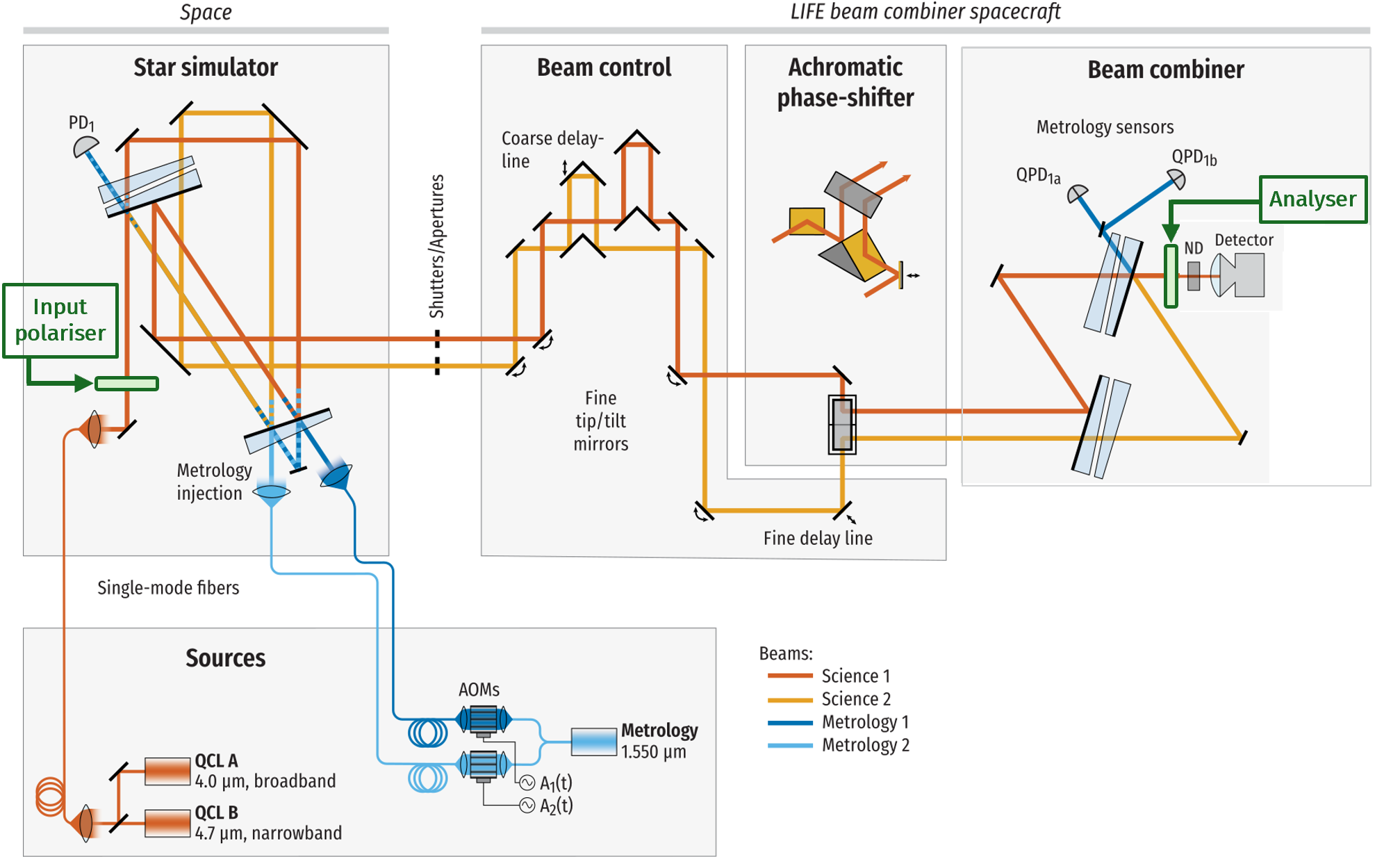}
    \caption{Schematic layout of the NICE bench during characterisation of the polarisation state of the two beams. The input polariser is used to propagate a purely $s$ or $p$ polarisation mode throughout the system. The analyser is used to measure the polarisation state after recombination.}
    \label{fig:bench_pol_characterisation}
\end{figure}

We use two polarisers: an input polariser placed after the fibre guiding the light from our different sources, and an analyser placed after the beam recombination and \textit{before} the spatial filter. \Cref{fig:bench_pol_characterisation} shows the locations of the two polarisers on the NICE layout. 
The two polarisers are then used to study the propagation of the $s$ and $p$ polarisation modes and the resulting states at the analyser.

Two measurements are recorded:
\begin{itemize}
    \item The maximum $I_\text{max}$ and minimum $I_\text{min}$ intensities measured by the camera when rotating the analyser,
    \item The angle $\theta$ on the polariser corresponding to $I_\text{min}$. We assume that the angle $\theta$ found at the minimum $I_\text{min}$ is exactly 90$^{\circ}$ from the angle found at the maximum $I_\text{max}$, which is consistent with our measurements. We use the $\theta$ measurement at the minimum instead of the maximum to obtain a higher precision.
\end{itemize}
A differential polarisation rotation $\delta\theta$ will generate a $\delta I_{sp}$ which will impact the null depth. Using the requirement on $\delta I_{sp}$ from \cref{eq:nice1_null-dynamic} gives a requirement of $\delta\theta<10'$.
The results of $I_\text{max}$ and $I_\text{min}$ are used to estimate the degree of circular polarisation of each beam; the differential of which relates to both $\delta \phi_{sp}$ and $\delta I_{sp}$. However, the direction of the circular polarisation cannot be estimated from this measurement. The use of a quarter waveplate would be necessary for this, and to confirm that $I_\text{min}$ indeed corresponds to circularly polarised light. We have obtained a Fresnel rhomb for this purpose (as it can be used as a broadband quarter waveplate), and will use it for future measurements. 

The second step is a measurement at the null depth for the different polarisation states using the setup in Fig.\,\ref{fig:optical_schematic}, with an input polariser after the source fibre, and the analyser placed \textit{after} the spatial filter.
Three measurements can also be obtained from this
\begin{itemize}
    \item The null depth for the different polarisation modes at the input polariser and the analyser (e.g., $N_{sp}$ would be the null depth found for input polarisation $s$ and analyser transmitting $p$),
    \item The phase difference when changing the analyser between $s$ and $p$ (i.e., phase difference between $N_{ss}$ and $N_{sp}$),
    \item The null depth without analyser (i.e., $N_s$ and $N_p$ for input polarisation $s$ and $p$, respectively).
\end{itemize}
The null depth measurements with the analyser can be compared to the requirement of $10^{-5}$, and used as a way to diagnose the main source of polarisation errors between the beams. The phase difference when changing the analyser gives an estimate for $\delta\phi_{sp}$. Finally, the null depth measurements without analyser give us the total null depth for one input polarisation mode. 
Because of the natural polarisation of the light at the output of the laser, null measurements were only made with the input polariser transmitting the $s$ mode; the $p$ mode was too faint. The future implementation of a half-wave plate will allow to rotate the polarisation state of the laser, and null measurements. Obtaining both $N_s$ and $N_p$ will give us an estimate of the polarisation agnostic null depth that NICE can reach.

We mention here hat the fibres we use are not polarisation-maintaining. As such, they will produce a random polarisation orientation as a function of wavelength, temperature and bend radius among other variables. While this is frustrating to deal with during characterisation, as the exact polarisation states in the system are not repeatable, this effect will work to ensure that the null is indeed polarisation agnostic as the polarisation may in fact rotate during a measurement. It is critical to note, however, that this effect only corresponds to the absolute polarisation state. The effects on the null are differential, and as long as the two beams are injected in the same manner into the fibre, this so-called ``polarisation scrambling'' should not influence the null beyond the asymmetries that need to be corrected in the first place. 

\subsubsection{Results} 

\begin{figure}
    \centering
    \includegraphics[width=0.75\linewidth]{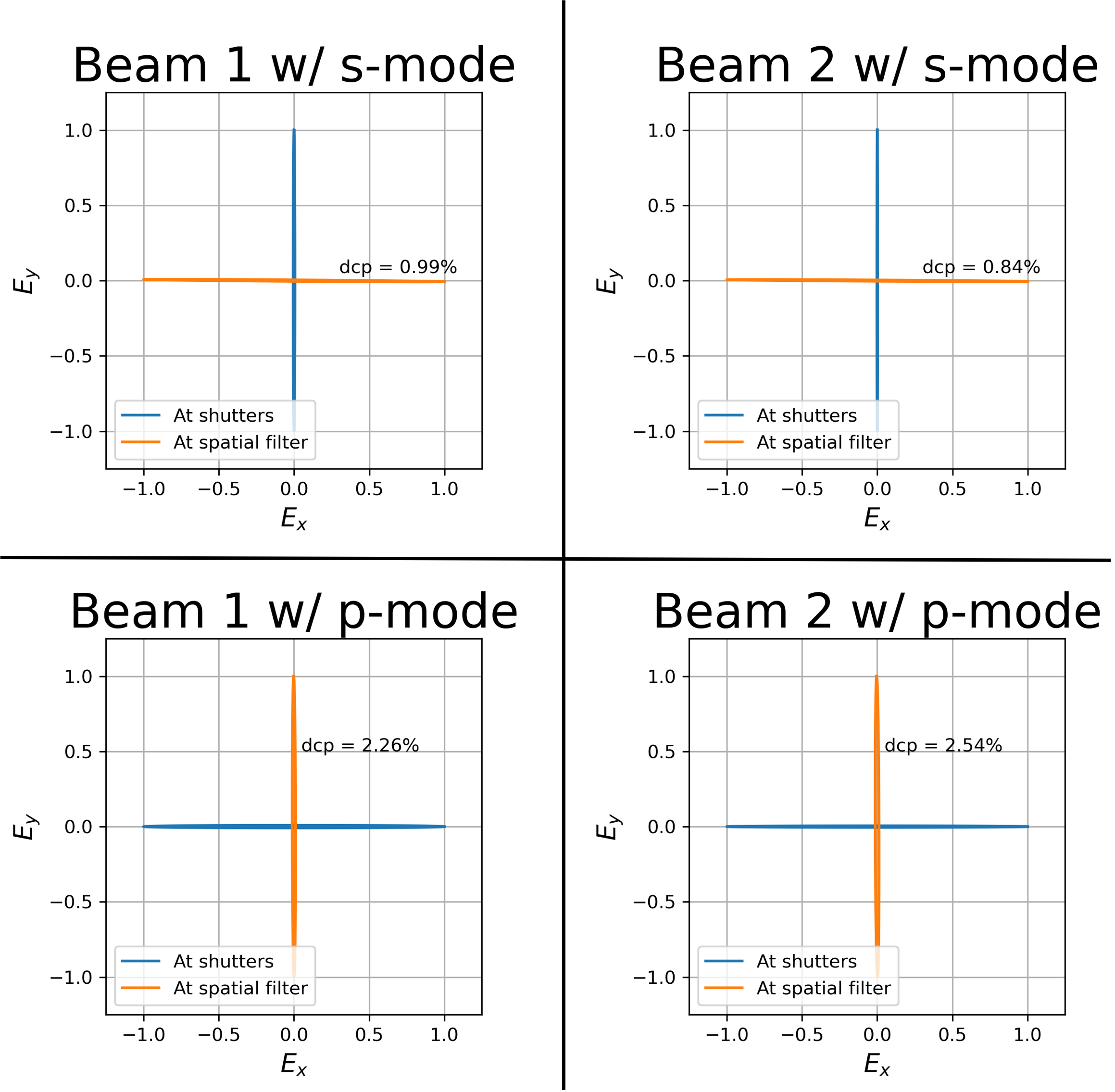}
    \caption{Results of the polarisation characterisation for the NICE test bench, for input polarisation in $s$ (top) and $p$ (bottom) modes. The results at the shutters position is indicated in blue, the results before the spatial filter are in orange. The degree of circular polarisation (dcp) found before the spatial filtre is also noted.}
    \label{fig:Pol_results}
\end{figure}

During the polarisation characterisation campaign, several measurements were performed at different locations of the bench. These measurements were able to locate the elements of the system that were generating differential polarisation effects between the beams. For example, a major error was generated by a differential out-of-plane roll angle of the two retroreflecting roof mirrors used in the delay lines. After its correction, we obtained the results summarised in Fig.\,\ref{fig:Pol_results} and in Table\,\ref{tab:Pol_results}. We find a typical difference in degree of circular polarisation ($\delta$dcp) between the two beams of 0.15$\pm$0.02\,\% and 0.28$\pm$0.02\,\% for the input $s$ and $p$ polarisation, respectively. The difference of polarisation angle ($\delta\theta$) is around 5'$\pm$5', mainly limited by the precision of our mount, and which is within the NICE requirements.

Table\,\ref{tab:Pol_results} also provides the results for $N_{ss}$, $N_{sp}$, and $N_{s}$. We find that the null is deeper when the analyser transmits the $p$ mode, but still higher than the best measurement obtained in Birbacher \& Hansen et al. (2026)\cite{Birbacher-Hansen-2026}. This is likely due to cladding mode leakage (see \cref{sec:spatial_filtering}), as the floor is on the same order as previous nulls that have not reached the requirement. 
$N_{ss}$ appears to be higher than $N_{sp}$ by a factor of five. Photometric imbalance in the $s$ mode is considerably higher than that of the $p$ mode, with null depth limitations on the order of $>10^{-3}$ and $>10^{-5}$ respectively, indicating that there is some differential polarisation effect present. A non-negligible portion of this deviation may also be due to a lower SNR, as the light is mainly $p$-polarised at the spatial filter; this would account for the fact that the predicted unbalance is higher than that measured for $N_{ss}$. Repetitions of these measurements with a diagonally polarised output beam is required to ensure a high confidence in the reported results.  

The requirement on ellipticity error $\delta$dcp still needs to be assessed. Since $\delta\theta$ is within requirements, the $\delta$dcp is probably the main reason to find $N_s > 10^{-5}$ and $N_{ss} > N_{sp}$, which likely comes from residual alignment error of the two beams in the beam control section. A study is on-going to assess if improvements on the alignment precision will be enough to reduce $\delta$dcp down to our requirement, or if a compensation mechanism should be implemented. What such a mechanism looks like is yet to be addressed, as while polarisation rotation can be controlled via something akin to a K-mirror, differential retardance will likely need to be controlled with a mechanism that also affects global alignment parameters (such as the delay-line roof mirror that was found to cause differential polarisation earlier).


\begin{table}[]
    \centering
    \caption{Summary of the differential degree of circular polarisation ($\delta$dcp) and differential polarisation angle ($\delta\theta$) between the two beams for $s$ and $p$ mode as input polarisation. The null depths obtained for $s$ input polarisation are also given, with the OPD difference $\delta\phi_{sp}$ obtained when rotating the analyser.}
    \begin{tabular}{c c c c c c c}
    \hline \hline
         & $\delta$dcp & $\delta\theta$ & $N_{ss}[10^{-5}]$ & $N_{sp}[10^{-5}]$ & $N_s[10^{-5}]$ & $\delta\phi_{sp}$ \\ \hline
        $s$-mode & 0.15$\pm$0.02\,\% & 5$\pm$5' & 58$\pm$10 & 9.8$\pm$0.9 & 49$\pm$1.4 & $\sim$16\,nm \\
        $p$-mode & 0.28$\pm$0.02\,\% & 5$\pm$5' & - & - & - & - \\ \hline
    \end{tabular}\\
    \label{tab:Pol_results}
\end{table}

\section{Cryogenic preparation}
\label{sec:cryo}

\begin{figure}
    \centering
    \includegraphics[width=0.8\linewidth]{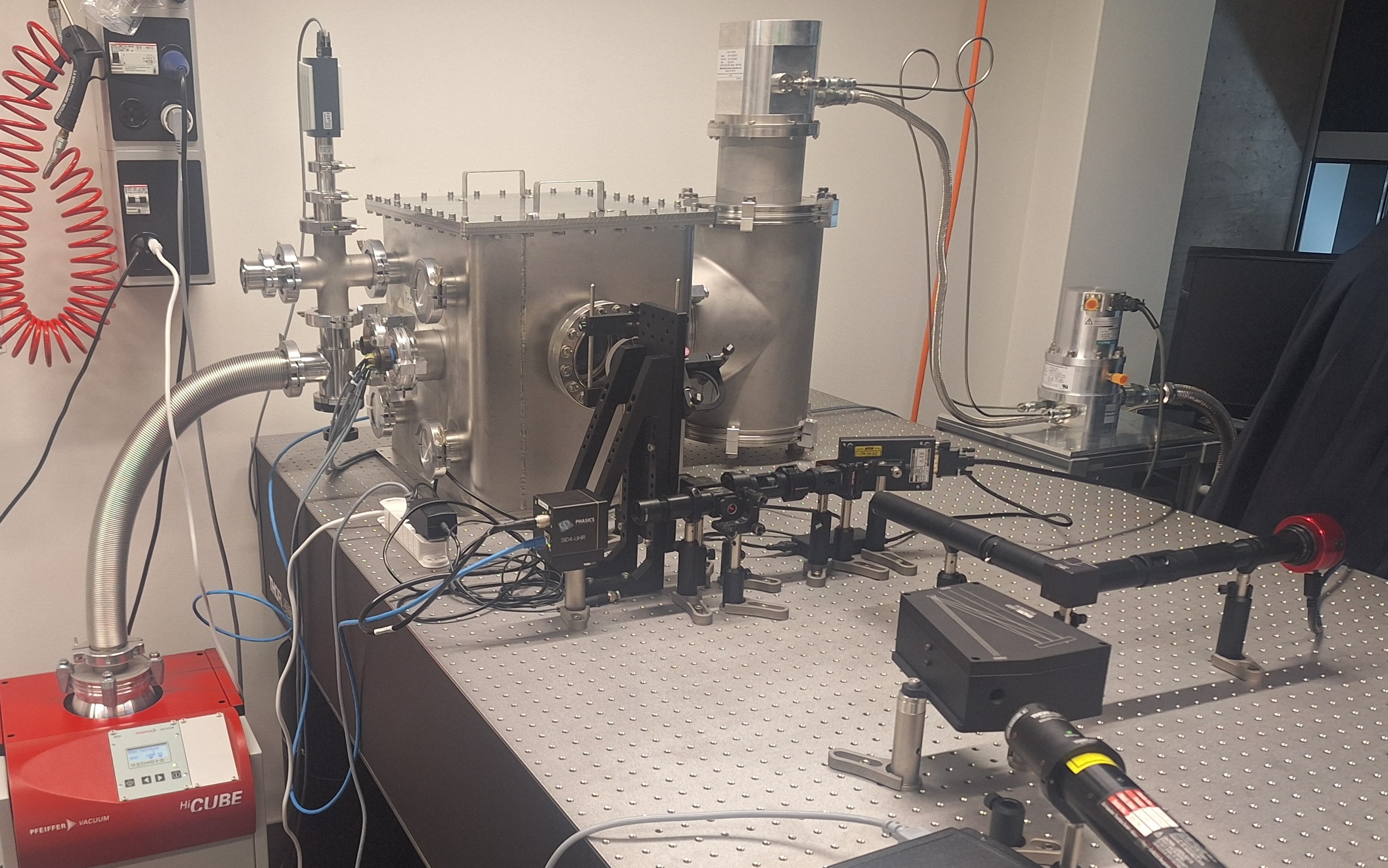}
    \caption{Photograph of the ``Ice Cube'' cryostat, along with the optical characterisation bench.}
    \label{fig:cryostat}
\end{figure}

Along with the ambient bench activities progressing the optical design, we have also begun work on preparing the setup for cryogenic operation. The key enabler in this task is the ``Ice Cube'' cryostat, a small modular cryostat designed to characterise component and subsystem level assemblies to monitor their deformation during cooldown to 15\,K. An image of the cryostat is shown in \cref{fig:cryostat}. More information on the cryostat, the optical measurement setup and some initial tests can be found in the complimentary proceeding by Hansen et al. (proc.  14154-97 \cite{Hansen_SPIE_2026}). 

In the near future, we plan to use this cryostat to measure a few different opto-mechanical mounting systems that align with the tolerances of the optical model discussed in \cref{sec:tolerancing}, and verify that any thermal deformation will either stay within the requirements, or else that  static alignment processes (such as shim) can compensate. We will also use this cryostat to test the functionality of our cryogenic actuators, as well as assess the capabilities of our SMFs and future fringe tracking photonic chips. Future upgrades will likely be needed to extend the measurements to more complex components such as deformable mirrors and detectors.

Finally, we have also begun work on designing the final NICE cryostat -- not to be confused with the Ice Cube mentioned above. This much larger vessel will contain the full NICE system, along with extra margin to allow future extensions and developments. The cooling apparatus and design will nevertheless be very similar to the Ice Cube and will implement the lessons learned with its operation, We anticipate a target temperature of about 15\,K, though the overall size is still to be confirmed. Whether the cryostat utilises a bench-type design, or some more unconventional integrated rail-type mounting system, is also being considered. We are aiming to finalise a design within the coming year.

\section{Future outlook and roadmap}
\label{sec:future}

\begin{figure}
    \centering
    \includegraphics[width=\linewidth]{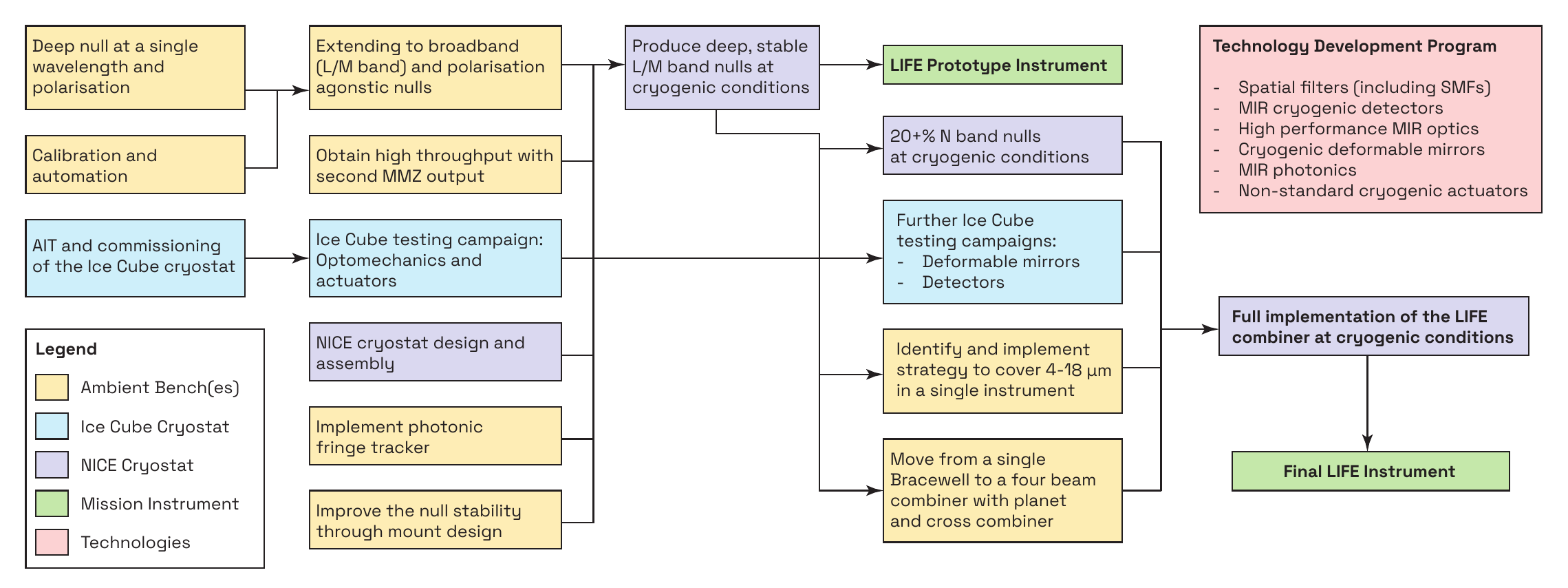}
    \caption{Roadmap for the NICE experiment, broken down into broad milestones and separated into the various benches (ambient, cryogenic etc.)}
    \label{fig:roadmap}
\end{figure}

The NICE experiment has made a lot of progress over the past few years, yet there is still much to do. In this final section, we detail the next steps for the project and the foreseen roadmap that takes NICE from a lab experiment towards a verification of the LIFE instrument. The roadmap is shown in \cref{fig:roadmap}.

\subsection{Towards the LIFE prototype mission}

A recently modified goal of the NICE experiment will be to support development of the LIFE prototype mission: a small connected element single-Bracewell nulling interferometer designed to demonstrate the LIFE measurement principle in orbit (Glauser et al., proc. 14148-64\cite{Glauser_SPIE_2026}). This mission is planned to work in the L- and M-bands, and the instrument concept is nearly identical to that of the current NICE ambient bench. As such, a first priority on the cryogenic side will be to emulate the current ambient bench at 15\,K, with the added inclusion of two MMZ outputs and a fringe tracker. 

To reach this goal, we identify a few milestones along the way:
\begin{enumerate}
    \item Identify and fix the issues on the ambient bench that are preventing the achromatic nulls. This will likely involve modifying our opto-mechanical design to use custom, integrated structures rather than off-the-shelf kinematic mounts.
    \item Implement an automated alignment and calibration routine to speed up iterations on the bench and increase stability.
    \item Work to obtain the two nulled outputs of the MMZ simultaneously, thereby doubling the throughput of the instrument.
    \item Fully commission the Ice Cube cryostat and undertake a measurement campaign to inform the cryogenic optomechanical design
    \item Design, procure and commission the larger NICE cryostat
    \item Manufacture and implement a photonic fringe tracker, nominally working in K-band.
\end{enumerate}

\subsection{Towards an even wider bandpass and the LIFE combiner}

Beyond the development of the prototype, and in parallel with those activities, we also need to focus on the wider LIFE instrument, its longer wavelengths and the octave of wavelength coverage. Along with implementing longer wavelength sources (e.g. a \ce{CO_2} laser and a blackbody) and replacing the optics/coatings as needed, there are two major upgrades to the experiment that are needed.

First, we must deal with the requirement of an octave of wavelength coverage. There are essentially two variations that are to be traded: either we split the wavelength coverage into multiple sub-wavelength channels, or we actively compensate the wavelength dependent error directly. For the former, this requires MIR dichroics, which currently do not have the best throughput or efficiency. An alternative is to spectrally disperse the light and use the equivalent of an image slicer (such as those used in Integral Field Spectroscopy) to separate into channels. For the latter, the simplest option is to implement an adaptive nuller and actuate the chromatic phase and amplitude directly. Of course, this requires the development of a cryogenic deformable mirror. One could even foresee coupling the two concepts together, using an adaptive nuller and then a slicer in series to ensure that the SMFs and beam splitter coatings are optimal over a smaller sub-bandpass. We point the reader to the proceeding of Birbacher et al. (proc. 14148-119 \cite{Birbacher_SPIE_2026}) for a longer discussion on this trade.

Secondly, NICE should also eventually move to a double Bracewell beam combiner. To facilitate this, the ``star simulator'' must be extended to four beams. This could be implemented either as a cascade of the current modified Sagnac splitters (thus resulting in a heavy loss of throughput), or equivalently as a so-called ``wavefront divider/separator'', which takes a single beam and samples multiple sub-apertures \cite{LeDuigou_2012,Garreau_2024}. This latter approach is conceptually simpler and also allows for an easier implementation of a faux planet signal: providing a second source at a slight tilt corresponding to the off-axis angle of the planet. The downside is that the intensity distribution across each sub-aperture will be neither Gaussian, nor top-hat, and each beam will have a differential polarisation across the pupil. Nevertheless, LIFE will experience similar problems in its optical design, so figuring out potential solutions (such as PIAA apodisation \cite{Guyon2003}; see Garreau et al., proc. 14148-120 \cite{Garreau_SPIE_2026}) is beneficial regardless.

The move to a double Bracewell combiner does not only mean duplicating the single Bracewell design, but also entails implementing the cross-combiner. This involves designing an achromatic, variable $\pm\pi/2$ phase shift to perform phase chopping, which provides new challenges in terms of cryogenic mechanisms. 

\subsection{Technology developments}

Along with system level design and verification, there are a number of component level technologies that need to be developed simultaneously. These are listed in the top corner of \cref{fig:roadmap}, and include:
\begin{enumerate}
    \item Spatial filters that exhibit high modal rejection at the level of $10^{-6}$, high throughput (ideally $>$50\% including coupling and propagation losses) and function over a wide wavelength range. They also must function at 15\,K.
    \item MIR cryogenic detectors. While NICE has detectors that function over L and M band, including a H2RG MCT detector for cryogenic operation. However, for longer wavelengths we lack options for detectors with low enough dark current. As such, developments into low noise, superconducting detectors such as MKIDs \cite{Ras-Vinke_2026} are critical.
    \item The development of high performance, broadband AR coatings and beam splitter coatings with minimal polarisation dependence. This reduces the requirements on the optomechanical alignment and actuators as described in \cref{sec:chromatic}.
    \item Cryogenic deformable mirrors. In the context of the adaptive nuller subsystem for explicit compensation of chromatic error, a key component is that of a DM that functions in a well-behaved way at 15\,K. This has been demonstrated before using a MEMS DM \cite{Takahashi_2017}, but more effort is needed to raise the TRL level and implement into the setup. Other DM technologies may also be suitable\cite{Huisman_2021}.
    \item MIR photonics - as discussed in Glauser et al. (proc. 14148-64\cite{Glauser_SPIE_2026}), photonic chips could potentially be used to replace many portions of the beam combiner, alleviating issues regarding relative alignment and greatly reducing mass and volume of the instrument. MIR platforms such as InGaAs, while showing considerable promise \cite{MontesinosBallester_2024}, nevertheless still need to demonstrate high throughput and deep nulling before a trade can be seriously considered. 
\end{enumerate}

\subsection{Conclusion}
Overall, NICE has made some great progress over the past year. While there is still a lot more work to do, the path forward and development roadmap is clear. With the LIFE prototype mission heading into a concept design stage in the coming months, we anticipate work on NICE to continue to ramp up, and look forward to pressing forward with the implementation of the full LIFE mission over the coming years. 

\comment{
\section*{Trash/possible appendix}
The measurements of $I_{min}$ and $I_{max}$ give us the degree of circular polarisation of the beam, assuming again that the beam is fully polarised and that $I_{min}$ is not due to unpolarised light. $I_{min}$ and $I_{max}$ can be used to calculate $E_p$ and $E_s$ depending on the input polarisation state. The degree of circular polarisation $dcp$ is obtained from
\begin{equation}
    dcp = -2\frac{E_p\times E_s}{E_p^2+E_s^2}.
\end{equation}

where we can relate the $s$ and $p$ amplitudes of the electric fields $E_{s}$ and $E_{p}$ to $\delta I$ and $\delta I_{sp}$
\begin{align}
    \delta I &= \frac{E_{1p}^2 + E_{1s}^2 - E_{2p}^2 - E_{2s}^2}{E_{1p}^2 + E_{1s}^2 + E_{2p}^2 + E_{2s}^2}, \\
    \delta I_{p} &= \frac{E_{1p}^2 - E_{2p}^2}{E_{1p}^2 + E_{2p}^2} \\
    \delta I_{s} &= \frac{E_{1s}^2 - E_{2s}^2}{E_{1s}^2 + E_{2s}^2}
\end{align}

The difference of $\theta$ between the beams is also related to the requirement on $\delta I_{sp}$. A rotation of beam 1 by $\theta$ corresponds to a change in its Jones vector as
\begin{equation}
    |1\rangle = 
    \begin{pmatrix}
    E_{1p}\cos(\theta)+E_{1s}\sin(\theta)e^{i\delta\phi_{1sp}} \\ E_{1p}\sin(\theta) + E_{1s}\cos(\theta) e^{i\delta\phi_{1sp}} 
    \end{pmatrix}.
\end{equation}
Assuming $|2\rangle$ do not rotate, one obtains
\begin{equation}
    \delta I_{sp} = \frac{\cos(\theta)^2-1}{\cos(\theta)^2+1} \underset{\theta \ll 1}{\sim} \frac{-\theta^2}{2},
\end{equation}
which gives a requirement of $\theta\leq7^{\circ}$.
}

\acknowledgments 
 
Part of this work has been carried out within the framework of the National Centre of Competence in Research PlanetS
supported by the Swiss National Science Foundation under grants 51NF40 182901 and 51NF40 205606. This work was also supported by the Swiss National Science Foundation (grant number 10004532) and by the Swiss State Secretariat for Education, Research and Innovation (SERI) / Swiss Space Office (SSO). This project was supported by Rudolf Bär via the ETH
Zurich Foundation. No AI tools were used in either the analysis or writing of this manuscript.
\bibliography{report} 
\bibliographystyle{spiebib2} 

\end{document}